# Ion-Specific Effects at the Surface of Water

Sanghamitra Sengupta[*†], Jan Versluis, and Huib J. Bakker[*†]
AMOLF, Science Park 104, 1098 XG, Amsterdam, The Netherlands
Corresponding author's email: s.sengupta@amolf.nl
and h.bakker@amolf.nl

**Abstract:**

We studied the interaction between salts and surfactants on the water surface using heterodyne-detected vibrational sum frequency generation (HD-VSFG) spectroscopy. We used sodium dodecyl sulfate (SDS) as a prototype surfactant system at 75 µM bulk concentration in water. The vibrational response of the OH band of near-surface oriented water molecules and the CH bands of the hydrophobic tails of the surfactant are measured. We observed a dramatic enhancement of the surface density of the negatively charged SDS (DS$^-$) within a narrow range of added salt concentrations. We demonstrated this increase is strongly ion-specific, and induced by the screening of the lateral Coulomb repulsion of the sulfate headgroups by the added cations, followed by strong hydrophobic interactions (hydrophobic collapse) when the DS$^-$ surface density reaches a critical value. For a solution of 75 µM SDS, the required concentrations of CsCl, KCl, and NaCl for this transition are 2, 5, and 10 mM, respectively.

The interfacial molecular framework of charged interfaces is embedded with a plethora of information. This information is vital for the understanding of several geochemical[1], and biophysical processes[2,3]. The sub-microscopic understanding of the properties of these interfaces also enables the smart design of optoelectronic devices that can be used for energy storage[4], and energy conversion[5,6]. A crucial role in the properties of charged interfaces is played by the presence of other charged species, i.e. ions, in the adjacent phases. The conformations and functionalities of macromolecular assemblies such as proteins[7,8], carbon nanotubes[9,10], and catalysts[11,12] are also heavily dependent on their interaction with ions and often can be controlled by varying their concentration and nature.

Ions play a pivotal role in tailoring the electrostatic energy landscape at the interface[2,13,14]. They differ strikingly in their physiochemical properties based on their ionic charge, radius[15], atomicity[8,16], etc. The Hoffmeister series is one of the concepts that aided the deciphering of ion-specific effects on biological and environmental systems[7,17], providing a means to predict so-called salting-in and salting-out processes. The series is constructed based on experimental observations and the mechanisms behind the predictions of this series are a topic of continuous exploration. There are many exceptions to the series that are as of yet poorly understood, particularly systems under biologically relevant conditions.[18] Low ion-strength regimes are quite important for biological systems but scarcely explored mostly due to lack of sufficiently selective and sensitive experimental tools. These leave us with a lot of unanswered but important questions about the structure and energetics of complex interfaces with undeniable importance.

A relatively new and powerful technique to investigate aqueous surface is heterodyne-detected vibrational sum frequency generation (HD-VSFG) spectroscopy. Over the last decade, several studies have been reported employing HD-VSFG to study the effects of ions on the properties of water covered with differently charged surfactants and biologically relevant macromolecules[19-25]. It was found that the orientation of the water molecules is determined by the net charge of the surfactant layer[19,20,22]. It was also observed that the addition of salts affects the degree of orientation, as a result of the screening of the electric field exerted by the charged surfactant layer. For positively charged surfactants, the effect of the added anions on the water signal strongly depends on the nature of the anion, following the Hofmeister series, while for negative surfactant layers the water signal was observed to depend only weakly on the nature of the added cation[22].

Here we present a study of the effect of adding chloride salts with different cations on the water signal and surfactant surface density for an aqueous solution of sodium dodecyl sulfate (SDS). SDS is a ubiquitous negatively charged surfactant that has been employed to mimic charged biological interfaces. Our work differs from previous studies in that we investigate the low-concentration regime down to a few millimolar of added salt, whereas in previous work salt concentrations of 0.5 M were used[22]. In this low-concentration regime, we observe very strong effects of the added salt on the water signal and surfactant surface density, that strongly depend on the nature of the cation.

We probe the response of the OH stretch vibration of oriented water molecules near the surface and the response of the CH stretch vibrations of the hydrophobic tail of the dodecyl sulfate (DS⁻) at the surface. We chose sodium chloride (NaCl), potassium chloride (KCl), and caesium chloride (CsCl), i.e. salts with monoatomic, univalent cations of group 1 of the periodic table that show a large difference in ionic radius. We studied the effect of the addition of these salts at concentrations ranging from the millimolar to molar regime.

In Figure 1 we present the extracted imaginary part of the HD-VSFG spectra of an aqueous solution of 75 µM DS$^-$ measured with different added concentrations of sodium chloride CsCl. The HD-VSFG spectra show the presence of a broad positive signal around 3250 cm$^{-1}$ representing the response of the OH vibration of oriented water molecules near the surface.[26] The positive sign of the water signal implies that the water molecules are pointing with their H atoms toward the surface, as a result of the net negative charge of the surface resulting from the presence of DS$^-$-surfactants.

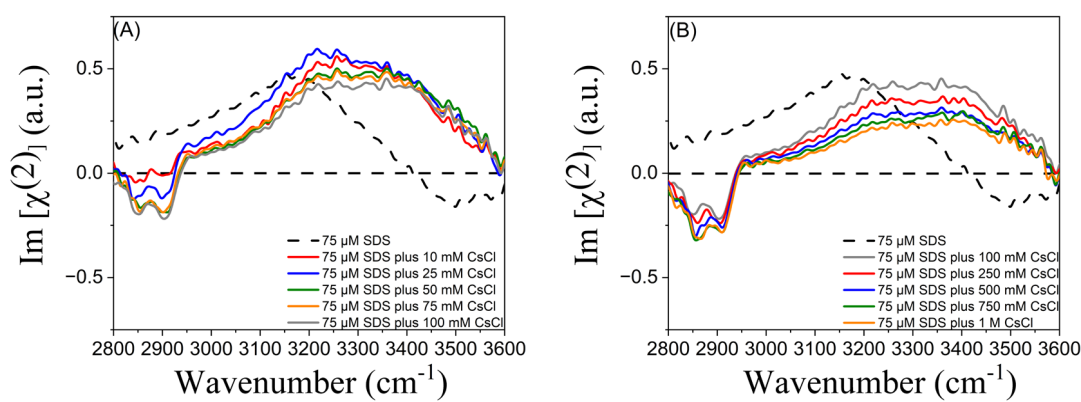

*Figure 1*: Heterodyne detected vibration sum-frequency generation (HD-VSFG) spectra of an aqueous solution of 75 µM SDS and different CsCl concentrations. Panel (a) shows the salt concentration varying from 0 mM to 100 mM, and panel (b) shows the salt concentration varying from 100 mM to 1 M.

We observe that for a solution containing only SDS (dashed curve, no added salt), the HD-VSFG signal has a phase distortion because the Debye screening length is ~30 nm, leading to signal contributions from deeper below the surface. For added salt concentrations larger than a few mM, this phase distortion becomes negligible because the Debye screening length gets <5 nm. The spectrum of the O-H vibrations of water close to a charged surfactant layer contains two contributions, i.e. the response of the bonded interface layer (BIL), representing water molecules that are directly hydrogen bonded to the surfactant headgroups, and the response of water molecules in the diffuse layer (DL) below the interface that are oriented due to the electric field exerted by the surfactant layer[27-29]. These two responses can have a different spectrum. It is seen in Figure 1 that for CsCl concentrations between 10 and 250 mM, the shape of the O-H stretch response of the spectrum does not change, only the amplitude changes. We thus conclude that at salt concentrations <250 mM, the response of the DL layer completely dominates over the BIL response. At salt concentrations >250 mM the BIL response will become relatively more important, because at these higher concentrations the DL response will become small due to very strong screening of the surface electric field by the ions. The BIL response will be dominantly formed by water molecules hydrogen-bonded to the sulfate groups of the DS$^-$ surfactant. These hydrogen bonds will likely be somewhat weaker than those between water molecules in the bulk, thus explaining the small blue shift of the spectrum.

For a pure SDS solution the HD-VSFG spectrum does not show a response in the frequency region of the CH stretch vibrations (2800 - 3000 cm$^{-1}$). This indicates that the surface density of DS$^-$ is low. In addition, for a disordered or loosely packed monolayer of DS$^-$, the resultant responses of the CH$_2$ and CH$_3$ stretch vibrations become small due to the random/sporadic orientation of the hydrocarbon tails,[30] leaving no trace of those bands in the HD-VSFG spectra. Upon adding CsCl, two negative bands located at 2850 cm$^{-1}$ (contribution from both CH$_2$ and CH$_3$ symmetric stretch) and 2920 cm$^{-1}$ (Fermi resonance between CH$_3$ symmetric stretch and bending over tone) start to appear and show a rapid intensity enhancement with increasing CsCl concentration. This indicates the formation of an ordered DS$^-$ monolayer. At a CsCl concentration of ~50 mM the signals of the CH vibrations saturate.

In Figure 2 we present the amplitude of the OH stretch vibrations of water **(measured at a sum-frequency corresponding to an infrared frequency of 3300 cm$^{-1}$**, Figure 2a) and the symmetric CH$_3$ stretch vibration (measured at a sum-frequency corresponding to an infrared frequency of 2850 cm$^{-1}$, Figure 2b) as a function of the added salt concentration for NaCl (black squares) and CsCl (circular red dots). Figure 2a shows that the amplitude of the OH vibrations of the water molecules, has a quite complex dependence on the added salt concentration. We observe three different regimes of water HD-VSFG response with increasing salt concentration: (i) an initial fast decrease at salt concentrations <10 mM, (ii) a rapid rise for concentration 10-100 mM, and (iii) a gradual decrease at concentrations >100 mM**.** The three different regimes are more clearly discernible in Figure 3. Interestingly, we see that the different regimes of the water HD-VSFG response correspond to much lower salt concentrations for CsCl than for NaCl. Figure 2b shows the rise of the amplitude of the CH signal induced by adding NaCl or CsCl. The rise of the CH stretching signal with salt concentrations is much faster for CsCl than for NaCl. For both salts, the amplitude of the CH bands saturates, indicating that at salt concentrations >100 mM the surface density of DS$^-$ and its ordering remain unaltered. The HD-VSFG spectra at all measured NaCl and CsCl concentrations are shown in Figures SI1 and SI2 of the Supplementary Information (SI).

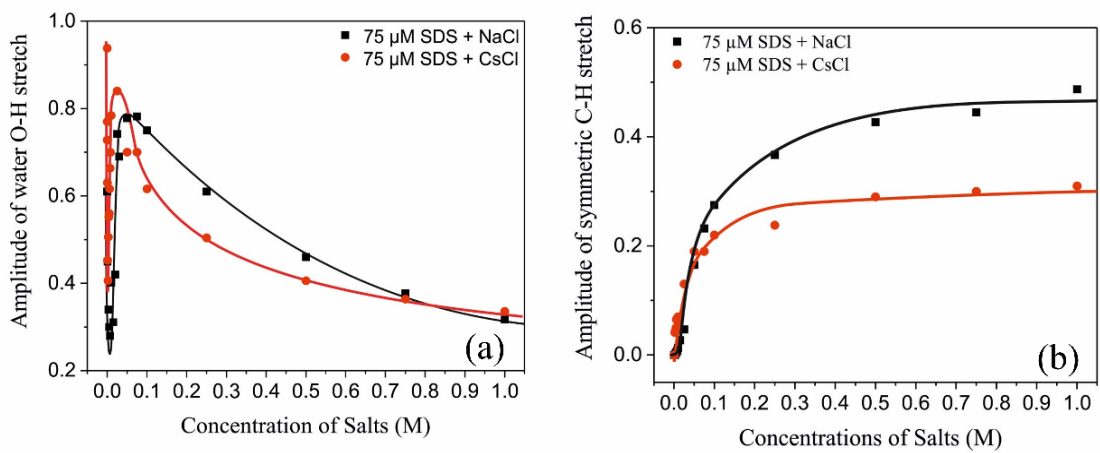

*Figure 2*: *The amplitude of the HD-VSFG signal of the water OH stretch vibrations (left panel, **measured at a sum-frequency corresponding to an infrared frequency of 3300 cm-$^1$**) and of the symmetric CH$_3$ and CH$_2$ stretch vibrations (right, **measured at a sum-frequency corresponding to an infrared frequency of 2850 cm-**) of an aqueous solution of 75 μM SDS as a function of the concentration of added NaCl (black solid squares) and CsCl (red solid circles). The symbols represent the experimental data, and the lines represent guides to the eye.*

The results of Figure 2 demonstrate that the interaction of DS$^-$ surfactants at the water surface and salts is highly salt-specific. To study this interaction further, we investigated the low-concentration regime in more detail, and also include KCl in the measurements. The choice of K$^+$ as a cation was motivated by its position in the periodic table and in the Hofmeister series (where K$^+$ seats in between Na$^+$ and Cs$^+$). The spectra measured for different concentrations of KCl and 75 μM SDS are shown in Figure SI 3.

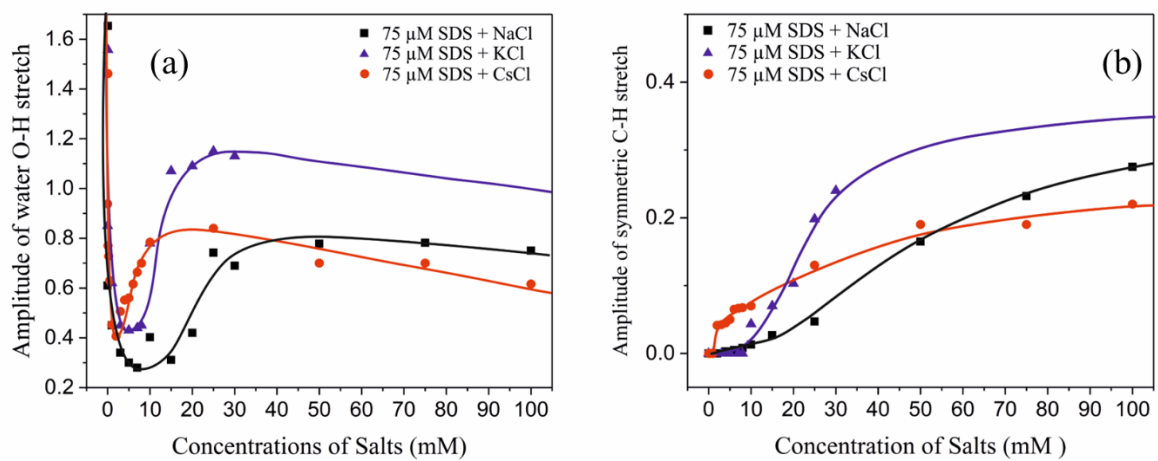

*Figure 3*: *The amplitude of the HD-VSFG signal of the water OH stretch vibrations (left panel) and of the symmetric CH$_3$ and CH$_2$ stretch vibrations (right) of an aqueous solution of 75 μM SDS as a function of the concentration of added NaCl (black solid squares), KCl (blue solid triangles) and CsCl (red solid circles). The symbols represent the experimental data and the lines represent guides to the eye. **The water OH stretch signals***

*are measured at a sum-frequency corresponding to an infrared frequency of 3300 cm$^{-1}$, the CH signals at a sum-frequency corresponding to an infrared frequency of 2850 cm$^{-1}$.*

In Figure 3a it is seen that the overall shape of the dependence of the amplitude of the OH stretch vibration is the same for all three studied salts, but that the corresponding concentration ranges strongly differ. For all three salts, the amplitude strongly decreases with salt concentration in the low concentration regime, but the concentration values at which the amplitude reaches a minimum strongly differ. For CsCl the minimum is attained at ~2 mM, for KCl at ~6 mM, and for NaCl at ~10 mM. After the minimum, the amplitude rises again, reaching a maximum at a very different concentration for the three salts. For CsCl the maximum is reached at ~18 mM, for KCl at ~27 mM, and for NaCl at ~40 mM. Figure 3b shows a similar strong difference in concentration dependence for the amplitude of the symmetric $CH_3$ and $CH_2$ vibrations for the three salts. For all three added salts, the amplitude shows a rise with initially fast increasing slope to then a smaller rise. The inflexion points of the signal increases are at very different concentration values for the three salts. For CsCl the inflexion point is at ~3 mM, for KCl at ~25 mM, and for NaCl at ~45 mM.

To investigate the potential role of the anion in affecting the water signal and the signal of the hydrophobic tails, we compare the results on NaCl with measurements of aqueous solutions of different concentrations of NaI and 75 μM SDS. The HD-VSFG spectra of the NaI solutions are presented in Figure SI4. Iodide (I$^-$) was chosen because it presents the highest contrast with Cl$^-$ in terms of ionic radius and charge density. We observed very similar results for the OH and CH responses for the NaI solutions as for solutions of NaCl at the same concentration. Hence, we conclude that the anion does not play a role and that the observed effects of the added salts on the DS$^-$ surfactant surface density and the response of water are fully determined by the nature and concentration of the cation.

To interpret the observations, we performed calculations with a model (see section Theoretical model) in which we combine the Poisson-Boltzmann equation with the Langmuir isotherm for adsorption of DS$^-$ to the surface. We extend the Poisson-Boltzmann equation with an energy term $\Delta G_s$ representing the surface propensity of the added cations. In the Langmuir isotherm, we include a term $\Delta G_{el}$ representing the electrostatic energy, which is proportional to the surface potential, and an energy term $\Delta G_W$, representing the favorable interaction of the hydrophobic tails of the DS$^-$ surfactants. At low surface densities the hydrophobic interaction will show a nonlinear dependence on the surface density, as the interaction between hydrophobic groups increases proportional to r$^{-6}$, with r the mutual distance of the hydrophobic groups. At higher surface densities $\Delta G_W$ will increase linearly with the surface density, as the DS$^-$ ions will form densely packed islands.

The Poisson-Boltzmann expression yields the surface potential as a function of the surface density, the ion concentration, and the energy term $\Delta G_s$, representing the cation surface propensity. The Langmuir isotherm yields the surface density in dependence of the surface potential and the surface density, because of the $\Delta G_W$ term, representing the hydrophobic interaction. The model thus comprises a set of coupled equations that are solved self-consistently for each added salt concentration, yielding the surface potential and the surface density at that concentration. We take the OH signal measured with HD-VSFG is proportional to the surface potential, as the observed OH signal is dominated by the response of the diffuse layer (DL), i.e. representing the degree of orientation of the water molecules in the near-surface region induced by the surface potential. We further assume that the CH signal is proportional to the surface density in the regime that the hydrophobic interaction $\Delta G_W$ increases linearly with the surface density, as this regime represents the response of the islands at the surface in which the hydrophobic tails of the DS$^-$ ions will be aligned and thus well-oriented.

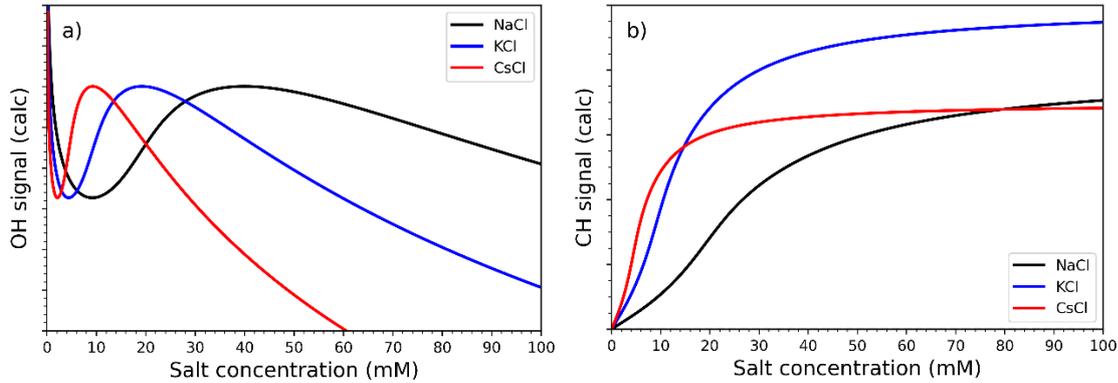

*Figure 4*: Calculated amplitude of the water OH stretch (left panel) and of the symmetric CH$_3$ and CH$_2$ stretch vibrations (right) of an aqueous solution of 75 μM SDS as a function of the concentration of added salts. The lines are calculated with the model described in the main text and the Theoretical model.

Figure 4 represents the calculated OH signal and the CH signal amplitudes as a function of the added salt concentration, obtained with the model. These calculations are performed with a DS$^-$ concentration of 75 μM, a surface adsorption energy of Na$^+$ of 0, of K$^+$ of 1.5 kJ/mol (0.6 $k_B$T per ion), and of Cs$^+$ of 3 kJ/mol (1.2 $k_B$T per ion), and a maximum favorable hydrophobic interaction energy of $1.8 \cdot 10^{-20}$ J (4.5 $k_B$T) per DS$^-$ surfactant. The results are in qualitative agreement with the data of Figure 3, showing a pronounced dip in the OH signal amplitude at 2.5, 6, and 10 mM of added salt for CsCl, KCl, and NaCl, respectively, and a sigmoidally shaped increase of the CH signal amplitude with salt concentration that becomes less steep in the order CsCl, KCl, and NaCl.

We find that the observations can be explained by a combination of three effects induced by the added cations: (i) electrostatic screening of the electric field pointing in the bulk resulting from the excess negative charge at the surface of the DS$^-$ surfactants, and (ii) lateral electrostatic screening of the Coulomb repulsion between the negatively charged sulfate headgroups of the DS$^-$ surfactants, (iii) van der Waals interaction between the hydrophobic tails of the DS$^-$ surfactants, The first effect leads to a reduction of the degree of orientation of the water molecules in the near-surface layer, and thus to a rapid initial decay of the water signal as a function of added salt concentration. The second effect leads to an increase in the surface density of DS$^-$ which increases the amount of oriented water. The third effect comes into play when the DS$^-$ -surfactants get at a smaller distance than a critical value. Then the short-range van der Waals interaction sets in, leading to a quite fast increase in surface density and ordering of the hydrophobic tails in a small range of added salt concentrations. The overall effect is that with increasing salt concentration the water signal first decreases due to effect (i), then increases due to effect (ii) and (iii). After these two effects saturated, establishing a near-complete DS$^-$ layer at the surface, the signal decreases again due to the remaining effect (i). For the CH signal, no signal is observed until a critical concentration is reached at which effects (ii) and (iii) become significant, inducing a strong increase in the surface density and an enhanced ordering of the hydrophobic tails. After that the increase of the CH signal slows down and approaches a maximum that is defined by the signal of a complete DS$^-$ layer at the surface.

We find that the efficiency of the different cations in enhancing the surface density of DS$^-$, increases in the order Na$^+$ to K$^+$ to Cs$^+$. This trend can be well explained by the increase in ionic radius going from Na$^+$ to K$^+$ to Cs$^+$. The experimentally observed ionic radii of the cations are 102 pm for Na$^+$, 138 pm for K$^+$, and 170 pm for Cs$^+$.[31] Due to its large ionic radius, the charge density of Cs$^+$ is low, and its interactions with water dipoles are not as favorable as for K$^+$ and Na$^+$. As a result, Cs$^+$ will not be well embedded in the bulk of the aqueous solution and will have a high surface propensity, thus being very effective in inducing effect (ii), i.e. the screening of the Coulomb repulsion between the negatively charged sulfate headgroups. The observed order in cations agrees quite well with the Hoffmeister series for cations[7,17].

Interestingly, in a previous study of the interaction of different cations with a surface layer of DS$^-$, little effect of the nature of the cation on the water signal was observed.[22] The reason for this difference with our observations is that in this latter study the signals were only measured for an added salt concentration of 0.5 M, corresponding to the regime where the surface density of DS$^-$ has saturated and the water signal is slowly decreasing with added salt concentration because of the enhanced screening of the electric field exerted into the bulk by the surfactant layer. For future studies, it would be of interest to perform HD-VSFG studies of the surface water signal for different types of surfactants, as a function of added ion concentration in the regime of low added salt concentrations, including systems that are known to form exceptions to the Hoffmeister series.

Our results show that the addition of simple salts to a solution containing SDS strongly affects the surface density of DS$^-$ ions. Similar strong combined effects of ions and SDS have been observed before in the denaturation of proteins[32,33], and in the separation of single-walled carbon nanotubes[9,34]. The results presented here indicate that these effects are likely caused by the screening of the Coulomb repulsion between the negatively charged sulfate headgroups by the added positive ions, thus strongly enhancing the density of DS$^-$ at the surface of proteins and carbon nanotubes. In fact, the ion-surfactant interactions highlighted in this study likely form the origin of the more generally observed combined effect of ions and surfactants on the conformation of proteins[7,8]

In conclusion, we studied the effect of the addition of NaCl, KCl, and CsCl on the surface density of dodecyl sulfate (DS$^-$) surfactant ions on the water surface, using heterodyne detected vibrational sum-frequency generation spectroscopy. We measure both the signal of the OH vibrations of water molecules close to the surface, and the CH signal of the hydrophobic tails of the DS$^-$ ions. We observe that the addition of these salts leads to an anomalous dependence of the water signal on the salt concentration, showing a strong increase of the water signal within a limited range of added salt concentration, 2-18 mM for CsCl, 6-27 mM for KCl, and 10-40 mM for NaCl. These increases reflect a strong increase in surface density in these concentration ranges, as is confirmed by the signal of the CH vibrations of the hydrophobic tails of the DS$^-$ ions. The increase in surface density is enabled by the screening of the mutual Coulomb repulsion of the negatively charged headgroups of the DS$^-$ ions, and a favorable interaction of their hydrophobic tails. The concentration region in which the water signal starts to increase is observed to be strongly dependent on the nature of the cation, which can be explained by the increase of surface propensity in the series $Na^+$, $K^+$, and $Cs^+$.

The present observation of a strong specific salt effect in the surface density of a charged surfactant will have implications for our understanding of the effect of ions like $Na^+$, $K^+,$ and $Ca^{2+}$ on the ratio of neutral, charged, and zwitterionic constituents of biological membranes. We hope that our results will stimulate future experimental and theoretical work in this direction. Our observations may also find industrial applications, as charged surfactants are widely used in chemical industrial processes, and it is demonstrated that their surface density can be strongly increased by adding salts.

**Acknowledgment:** This work is funded by the EU Horizon 2020 project called SoFiA (Soap film-based Artificial photosynthesis) (grant agreement ID: 828838). The authors would like to thank all SoFiA project colleagues for insightful discussions.

**Author Contribution:**
S.S.G. and H.J.B. had planned the project outline. S.S.G. had prepared all the solutions conducted all the experiments and did the data analysis. J.V. and H.J.B have done the theoretical calculations and prepared the plots. S.S.G. and H.J.B. both prepared the figures and contributed to the manuscript text.

**Competing Interest:**
The authors declare no competing interest.
Please contact h.bakker@amolf.nl for any further correspondence and materials to be requested.

**Experimental Materials:**

The materials that are used in the experiments mentioned in this research are listed below. The SDS was brought from Sigma Aldrich. For each experiment and follow-up repetition, a fresh 10 mM solution was made from a 1M SDS solution using the serial dilution method. Using a serial dilution method a final 20 ml 75 µM SDS solution was prepared from the 1 mM solution. All solutions were homogenized for 1 minute. The salts were brought from Sigma Aldrich (purity > 99.99 %) and they were stored in the glove box under dry and nitrogen airflow. A 1 M or 0.5 M salt solutions were prepared freshly on the day of the measurements and the required concentration was made using the serial dilution method. Special restrictions were taken in case of the KCl salts due to the hygroscopic nature of the KCl salts and the stock solution was prepared inside the glovebox to avoid any degradation.

**Experimental Methods:**

Heterodyne-Detected Sum-Frequency Generation spectroscopy (HD-VSFG) is an upgraded version of the traditional vibrational Sum-Frequency Generation Spectroscopy[35-38] (VSFG) that allows to obtain the real and imaginary part of the second-order susceptibility value of the surface-specific vibrations from the experimental spectra. The details of the experimental setup and the theory of the HD-VSFG technique can be found elsewhere.[39,40]. In VSFG, a mid-infrared (mid-IR) pulse ($\omega_{IR}$) of resonant frequency of the vibrations of the surface attached molecules and a visible pulse ($\omega_{VIS}$) overlap with each other at the same space and time to generate a new blue shifted light ($\omega_{SFG}$) at the sum of the two incident IR and visible frequencies, where $\omega_{SFG} = \omega_{IR} + \omega_{VIS}$. VSFG becomes only non-zero in non-centrosymmetric media under the electric dipole approximation. So, the non-symmetry imposed by the interface makes the second-order susceptibility interface specific. The VSFG electric field can be represented as $E_{SFG} \propto \chi^{(2)} E_{VIS} E_{IR}$ where $\chi^{(2)}$ is the second-ordered non-linear susceptibility and the $E$ terms represent the electric fields of the infrared and visible pulses respectively. The frequency dependence of $\chi^{(2)}$ on the frequency $\omega_{IR}$ can be expressed as follows:

$$\chi^{(2)} (\omega_{SFG} = \omega_{IR} + \omega_{VIS}) = A_{NR} + \sum_{i=0}^{n} \frac{A_n}{\omega_n - \omega_{IR} - i\Gamma_n}$$

where $\omega_n, A_n$, and $\Gamma_n$ are the frequency, amplitude, and damping constant of the $n^{th}$ vibrational mode and $A_{NR}$ is the non-resonant background. In HD-VSFG the electric field is phase-sensitively detected by interfering with the field at $\omega_{SFG}$ generated by the sample with that of a local oscillator.

An additional advantage of the phase-resolved technique over conventional intensity measurements is that it provides direct information on the orientation of the transition dipole moments of the vibrations at the surface. This is a great advantage over traditional VSFG as HD-VSFG delivers much faster knowledge of the molecular orientation at the interface.

All measurements reported in this manuscript is measured with a homebuilt HD-VSFG spectrometer[41,42]. A regenerative Ti: sapphire amplifier (Coherent legend) generates 800 nm (3 mJ, 35 fs) pulses at a repetition rate of 1 kHz. The pulses are divided into two parts to generate the mid-IR and the visible pulses, respectively. A part of 2 mJ is sent to a tuneable homebuilt optical parametric amplifier (OPA) and difference-frequency generator stage. The generated IR pulses have a frequency spectrum centered at 3000 cm$^{-1}$ and an energy of 8 µJ per pulse sent to the sample. The other part of 1 mJ of the fundamental beam is used to generate the visible beam (800 nm) using an etalon. In the HD-VSFG technique, we collect VSFG light from both the sample and a local oscillator (gold). The $\omega_{SFG}$ beam of the local oscillator is delayed to the $\omega_{SFG}$ beam of the sample by passing the local oscillator $\omega_{SFG}$ beam through a 1 mm thick silica plate. After passing the two $\omega_{SFG}$ beams through a monochromator, the interference pattern of the two beams is collected with an electron-multiplied charged–coupled device (EMCCD-Andor Technologies). Using Fourier transformation, filtering, and back Fourier transformation, both the imaginary and the real parts of the sample HD-VSFG electric field are obtained. This electric field is divided by the HD-VSFG electric field generated by a reference z-cut quartz crystal, to correct the spectra for the spectral intensity profile of the infrared pulse. All spectra presented in this article are an average of three individual spectrum, typically collected for 90 seconds under constant nitrogen flow. All presented data are measured with an SSP polarization configuration, where the indexing order of the beams is SFG, visible, and IR.

**Theoretical model:**

The starting point is the one-dimensional Poisson-Boltzmann equation that is modified by adding a 'depth (z)- depending' free energy $\Delta G_{s,i} e^{\frac{-z}{l_{s,i}}}$ contribution factor to the electrostatic energy – the term 'i' is used for indexing the ions. $\Delta G_{s,i}$ is positive when the ion 'i' has an enhanced surface density and negative when the ion 'i' has a decreased surface density. $l_{s,i}$ represents the characteristics 'length scale' of the surface propensity.

$$\frac{d^2\emptyset}{d_{z^2}} = \sum_i \frac{c_i e}{\varepsilon_0 \varepsilon} \left[ (-1)^{k_i} e^{((-1)^{k_i} e\emptyset(z) + \Delta G_{s,i} e^{\frac{-z}{l_{s,i}}})/k_B T} \right] \quad (1)$$

Where $\emptyset$ is the electric potential, $c_i$ is the concentration of each monovalent ion 'i' present in solution, $e$ is the elementary charge, $\varepsilon_0$ the dielectric permittivity of the vacuum, $\varepsilon$ the relative permittivity, $k_B$ is the Boltzmann's constant, and T is the temperature. The value of $k_i$ equals 1 for cations and 0 for anions. We multiply both sides with $2*\frac{d\emptyset}{dz}$ and the equations become like following:

$$2 * \frac{d\emptyset}{dz}\frac{d^2\emptyset}{dz^2} = 2 * \frac{d\emptyset}{dz} \sum_i \frac{c_i e}{\varepsilon_0 \varepsilon} \left[ (-1)^{k_i} e^{((-1)^{k_i} e\emptyset(z) + \Delta G_{s,i} e^{\frac{-z}{l_{s,i}}})/k_B T} \right] \quad (2)$$

The left side is equal to $\frac{d}{dz}(\frac{d\emptyset}{dz})^2$. Substitution and integration yields:

$$\int \frac{d}{dz}(\frac{d\emptyset}{dz})^2 dz = \sum_i \frac{2c_i e}{\varepsilon_0 \varepsilon} \int \frac{d\emptyset}{dz} \left[ (-1)^{k_i} e^{((-1)^{k_i} e\emptyset(z) + \Delta G_{s,i} e^{\frac{-z}{l_{s,i}}})/k_B T} \right] dz \quad (3)$$

With the result:

$$(\frac{d\emptyset}{dz})^2 = \sum_i \frac{2c_i k_B T}{\varepsilon_0 \varepsilon} \left[ e^{((-1)^{k_i} e\emptyset(z) + \Delta G_{s,i} e^{\frac{-z}{l_{s,i}}})/k_B T} \right] + C_1 \quad (4)$$

Where $C_1$ is an integration constant. The right-hand side has been integrated with respect to $\emptyset$. For large z, $\emptyset$, and $\frac{d\emptyset}{dz}$ becomes 0, yielding $C_1 = -2\sum_i \frac{c_i e}{\varepsilon_0 \varepsilon}$, hence we obtain the following equation:

$$\frac{d\emptyset}{dz} = -\sqrt{\sum_i \frac{2c_i k_B T}{\varepsilon_0 \varepsilon} \left[ e^{((-1)^{k_i} e\emptyset(z) + \Delta G_{s,i} e^{\frac{-z}{l_{s,i}}})/k_B T} - 1 \right]} \quad (5)$$

The negative sign at the beginning of the equation means that $\emptyset$ decreases for a positive potential with increasing distance, meaning that for $\emptyset > 0$, $\frac{d\emptyset}{dz} < 0$. Furthermore,

$$\sigma = \int_0^\infty \rho_e dz = \varepsilon_0 \varepsilon \int \frac{d^2\emptyset}{z^2} = -\varepsilon_0 \varepsilon \frac{d\emptyset}{dz}\Big|_0 \quad (6)$$

Substituting equation (5) in the above equation we get the following value

$$\sigma = \sqrt{\sum_i 2c_i k_B T \varepsilon_0 \varepsilon \left[ e^{((-1)^{k_i} e\emptyset(0) + \Delta G_{s,i})/k_B T} - 1 \right]} \quad (7)$$

Which provides a direct relation between $\sigma$ and $\emptyset(0)$. In the case that In the case that $\Delta G_{s,i} \equiv 0$ for all ions $i$, meaning that there is no enhanced or diminished adsorption to the surface, equation (7) reduces to the well-known Grahame equation.

In case of the SDS as in our experiments, the anion that gets accumulated at the interface is DS⁻ and yielding a surface charge density as the following:

$$\sigma = \frac{e\theta_{DS^-}}{a} \quad (8)$$

With $\theta_{DS^-}$ the surface fraction of surfactant ions DS⁻ and 'a' is the surface taken by a single DS⁻ moiety. The $\theta_{DS^-}$ can be then expressed as following:

$$\theta_{DS^-} = \frac{K_{L_{DS^-}} c_{DS^-}}{1+K_{L_{DS^-}} c_{DS^-}} \quad (9)$$

Where,

$$K_{L_{DS^-}} = K^0_{L_{DS^-}} e^{(\Delta G_{el} + \Delta G_{W_{DS^-}})/k_B T} \quad (10)$$

where $K^0_{L_{DS^-}}$ is the equilibrium constant for the adsorption in the absence of a surface potential and specific interactions between the DS⁻ ions. The $K_{L_{DS^-}}$ differs from $K^0_{L_{DS^-}}$ due to the two free energy contributions, which are the electrostatic energy in nature:

$$\Delta G_{el} = e\emptyset (0) \quad (11)$$

resulting from the surface potential, which follows from the above equation between $\sigma$ and $\emptyset(0)$, and the free energy $\Delta G_{W_{DS^-}}$ associated with the van der Waals interaction of the hydrophobic tails of the DS⁻ ions. The latter term will be strongly dependent on the surface occupation fraction $\theta_{DS^-}$, and will increase with $\theta^3_{DS^-}$ at low $\theta_{DS^-}$ values, as the attractive hydrophobic interaction is proportional to $r^{-6}$ with 'r' being the mutual distance. At higher $\theta_{DS^-}$ values and at higher salt concentrations, the DS⁻ ions will accumulate in islands and the $\Delta G_{W_{DS^-}}$ will increase linearly with $\theta_{DS^-}$. Defining $\Delta G_{W_{DS^-}} = \Delta G^0_{W_{DS^-}}$ for $\theta_{DS^-} = 1$, then we obtain:

$$\Delta G_{W_{DS^-}} = \Delta G^0_{W_{DS^-}} F(\theta_{DS^-}) \quad (12)$$

with $F(\theta_{DS^-})$ a function that describes the transition from third-order dependence on the $\theta_{DS^-}$ to a linear dependence on the $\theta_{DS^-}$. $F(\theta_{DS^-})$ approaches 1 when $\theta_{DS^-} = 1$.

The equations 7 to 12 can be independently solved for a given bulk concentration $c_{DS^-}$, the different monovalent ion concentration $c_i$ present in the solution, and the free energy terms $\Delta G_{s,i}$ and $\Delta G^0_{W_{DS^-}}$. The latter two parameters are fitted to the experimental data, and we only included a non-zero $\Delta G_{s,i}$ for the added cation. The solution yields the surface occupation fraction $\theta_{DS^-}$, the surface potential $\emptyset(0)$, a der Waals interaction energy $\Delta G^0_{W_{DS^-}}$ and the surface propensity energy $\Delta G_{s,i}$ of the added cation.

**Acknowledgments:**

This work is funded by the EU Horizon 2020 project called SoFiA (Soap film-based Artificial photosynthesis) (grant agreement ID: 828838). The authors would like to thank Jan Versluis and SoFiA colleagues for the insightful discussions.

**Author Contribution:**

S.S.G. and H.J.B. had planned the project outline. S.S.G. had prepared all the solutions conducted all the experiments and did the data analysis. S.S.G. and H.J.B. both prepared the figures and contributed to the manuscript text.

**Competing Interest:**

The authors declare no competing interest.

Please contact s.sengupta@amolf.nl for any further correspondence and materials to be requested.